\documentstyle[prl,aps,preprint]{revtex}

\let\ssection=\section
\renewcommand{\section}{\setcounter{equation}{0}\ssection}

\begin{document}
\draft
\title{Vacuum Einstein equations in terms of curvature forms}
\author{Yu.N. Obukhov$^{1}$\footnote{On leave from: Department of 
Theoretical Physics, Moscow State University, 117234 Moscow, Russia}
and S.I. Tertychniy$^2$}
\address{$^1$Institute for Theoretical Physics, University of Cologne,
 D--50923 K\"oln, Germany}
\address{$^2$National Research Institute for Physical, Technical and 
Radio-Technical Measurements (VNIIFTRI), 141570 Mendeleevo, Russia}
\maketitle

\begin{abstract}
\noindent
A closed explicit representation of the vacuum Einstein equations
in terms of  components of curvature 2-forms is given. The discussion
is restricted to the case of non-vanishing cubic invariant of  conformal
curvature spinor. The complete set of algebraic and differential
identities connecting  particular equations is presented and their
consistency conditions are analyzed.
\end{abstract}
\bigskip\bigskip\bigskip
{\bf Short title}: {\it Einstein equations in $S$-forms}
\bigskip\bigskip
\section{Introduction}

As it is well known, in the standard formulation of general relativity
theory the space-time metric is considered as a fundamental gravitational
field variable while the connection and curvature are in a sense secondary
and are derived from it. All the three are physically significant objects: 
the metric defines causal space-time 
structure and determines time and length scales, the curvature distinguishes
 true gravitational effects from  purely inertial ones, while the 
connection encompasses the possibility of treating gravity as a gauge 
theory which is important, in particular, in attempts of finding 
unification with the other types of interaction. Quite remarkably,
the beautiful and clear mathematical formalism of  classical general 
relativity allows for several alternative dynamical descriptions of the
same physical system. 

Historically, the components of the metric tensor (in a specific gauge) were 
first used as phase space variables (``gravitational potentials") of the 
gravitational field (e.g., ADM approach, \cite{adm}). Closely related are 
the approaches using the (pseudo-) orthonormal tetrads or similar algebraic 
structures.

A more recent Ashtekar's approach \cite{ash} exploits the connection as 
a source of phase space variables, and a number of concrete realizations 
of such an idea have been developed (see \cite{giu,pel} and references 
therein).

There are, however, some other choices of  a basic
structure describing a state of the gravitational field. Recently a 
considerable attention has been attracted to the formulations in which the 
self-dual two-forms play a central role \cite{cdj1,sam,hft,cdj2,cdj3,mh} 
(see a review and some further references in \cite{rev}). It is certainly 
worthwhile to mention an important contribution to this subject by 
Pleba\'nski \cite{p1,p2} (cf. also \cite{isr}) who investigated relations
between the metric and special bases of self- and anti-self-dual complex
2-forms and proved a number of statements which underlie many of the modern 
developments \cite{cdj1,sam,hft,cdj2,cdj3,mh}. In particular, the first
order chiral gravitational action for vacuum general relativity was 
introduced in \cite{p1}. 

Additional interest to the alternative gravitational field variables 
formulations is attracted by  recent studies of formal similarities and
exact mappings between  Yang-Mills gauge theories of internal symmetry
groups and  gauge gravity models \cite{free1,lun,moh,sij,free2}. These
approaches proved to be useful, in particular, for the problems of 
constructing new exact solutions in gravity and  Yang-Mills gauge 
theories. The relations  between  gravitational instantons and  Yang-Mills
gauge fields are discussed in \cite{cdj4}, while  new singular $SU(2)$ 
spherically and cylindrically symmetric solutions with  confining
properties are reported in \cite{conf}.

In the present work, we investigate the case when the space-time curvature 
is considered as a primary characteristic of the gravitational field. 
Previously, there were attempts (using  different, mainly purely 
algebraic methods) to derive  space-time metric in terms of  
curvature \cite{ihrig,hall,que}, or in terms of  connection \cite{sch}. 
In the so-called purely connection formulation of  general relativity 
 \cite{cdj1,cdj3} the metric is effectively eliminated in favour of 
connection. Our aim is to go one step further and to eliminate both the 
metric and connection, leaving only the curvature components as  
fundamental variables. We do not address at the moment such problems as 
a construction of an action, Hamiltonian, momenta etc, leaving this for 
 further study. As a first necessary step, the very possibility of 
deriving a closed expression for the vacuum Einstein field equations 
directly in terms of the curvature {\it alone} is demonstrated in 
an explicit form. This is the main new result reported in our paper.

The structure of the paper is as follows. After explaining some basic
notations and conventions in section 2, in section 3 we introduce the
notion of $S$-forms and discuss the properties of these objects which 
play a central role in our approach. Our new observations are formulated 
in the proposition 6 and corollaries 7,8. These technical results 
essentially underlie all further derivations in the paper. In section 4 
we show how one can construct $S$-forms from the curvature $\Omega$-forms. 
The crucial importance of $S$-forms in the description of a
four-dimensional Lorentzian geometry is outlined in section 5. Section 6
contains our main result, the theorem 11, which gives a self-consistent
formulation of the vacuum Einstein equations in terms of the curvature
two-forms. The mutual relations between  different subsets of the 
resulting algebraic-differential system of field equations are found
in section 7. Our conclusions are summarised in section 8.

\section{Preliminaries and notations}

We shall consider a complexified tangent space over some point of a 
4-dimensional space-time $M$. The Greek indices ${\scriptstyle \alpha,
\beta}$, \dots{} are running over the set \{1,2,3,4\} enumerating the 
local space-time coordinates. The capital Latin (spinor) indices, undotted
${\scriptstyle A,B}$,\dots{} and dotted ${\scriptstyle{\dot A},{\dot B}}$,
\dots, run over the 2-element sets $\{0,1\}$ and $\{\dot 0, \dot 1\}$, 
respectively.

Further, $\varepsilon^{\alpha\beta\mu\nu}, \varepsilon_{\alpha\beta\mu\nu}$
and $\epsilon^{AB}, \epsilon_{AB}, \epsilon^{\dot{A}\dot{B}}, 
\epsilon_{\dot{A}\dot{B}}$ are the standard Levi-Civita symbols, 
4-dimensional and 2-dimensional, respectively. Spinorial $\epsilon$'s are 
used for  lowering and  raising of spinor indices in accordance with
the rules $\iota^{B}=\iota_{A}\ \epsilon^{AB}, \iota_{A} = \epsilon_{AB}
\ \iota^{B}, \iota^{\dot B}=\iota_{\dot A}\ \epsilon^{\dot{A}\dot{B}},
\iota_{\dot A}=\epsilon_{\dot{A}\dot{B}}\ \iota^{\dot B}$
(i.e. $\delta^B_A=\epsilon^{B\cdot}_{\cdot A}, \delta^{\dot B}_{\dot A}
=\epsilon^{\dot{B}\cdot}_{\cdot\dot{A}}$).

Finally, we recall the general definition of Hodge dual and of the 
so-called $\eta$-basis of the exterior algebra over $M$. Let us denote a 
coframe 1--form $\vartheta^{a}$ (which forms an orthonormal basis of the 
(complexified) cotangent space $\Lambda^{1}$) and its dual frame $e_{a}$, 
such that $e_{a}{\_\!\_\kern -0.2ex\raise 0.4ex\hbox{$\scriptstyle |$}}
\vartheta^{b}=\delta^{b}_{a}$. Let the volume 
4--form be $\eta =\vartheta^{1}\wedge\vartheta^{2}\wedge\vartheta^{3}
\wedge\vartheta^{4}$. Then the $\eta$-basis and the Hodge duals are 
defined as follows:
\begin{equation}
\eta_{a}=e_{a}{\_\!\_\kern -0.2ex\raise 0.4ex\hbox{$\scriptstyle |$}}
\eta = \ast\vartheta_{a},\label{3f}
\end{equation}
\begin{equation}
\eta_{ab}=e_{b}{\_\!\_\kern -0.2ex\raise 0.4ex\hbox{$\scriptstyle |$}}
\eta_a =\ast(\vartheta_{a}\wedge\vartheta_{b}),
\label{2f}
\end{equation}
\begin{equation}
\eta_{abc}=e_{c}{\_\!\_\kern -0.2ex\raise 0.4ex\hbox{$\scriptstyle |$}}
\eta_{ab} = \ast(\vartheta_{a}\wedge\vartheta_{b}
\wedge\vartheta_{c}),\label{1f}
\end{equation}
\begin{equation}
\eta_{abcd}=e_{d}{\_\!\_\kern -0.2ex\raise 0.4ex\hbox{$\scriptstyle |$}}
\eta_{abc} = \ast(\vartheta_{a}\wedge\vartheta_{b}
\wedge\vartheta_{c}\wedge\vartheta_{d}).\label{0f}
\end{equation}

\section{Algebra of $S$-forms}
\label{BL}

Many of the results presented in this section seem to be well known, however 
we find it useful to collect all the facts together and summarize them in a 
lucid way  using the  convenient formalism of  spinor--valued 
(complex) exterior forms on $M$. The main result of this section is 
formulated in the proposition 6.

Let us start our discussion with introducing a fundamental object -- a 
triad of complex 2-forms $S_{AB}=S_{AB\mu\nu}dx^{\mu}\wedge dx^{\nu}$ which 
are labeled by a pair of spinor indices ${\scriptstyle A,B}$. This object 
is assumed to be symmetric, $S_{AB}=S_{BA}$, thus indeed representing three 
forms denoted by $S_{00}, S_{01}=S_{10}, S_{11}$.

Their properties are as follows. These forms are assumed to be 
non-degenerate in the sense that
\begin{equation}
S^{KL}\wedge S_{KL}\neq 0,\label{nodeg}
\end{equation}
and form a complete set satisfying
\begin{equation}
S^{AB}\wedge S_{CD}= \textstyle{1\over 3}\delta^{A}_{(C}\,\delta^{B}_{D)}\;
S^{KL}\wedge S_{KL}.\label{dcompletc}
\end{equation}

In fact, the non-degeneracy condition (\ref{nodeg}) introduces on the 
the space-time manifold a non-trivial volume 4-form which we denote
\begin{equation}
\eta:=-\textstyle{1\over 3}S^{KL}\wedge S_{KL}.\label{eta}
\end{equation}
The sign is related to the orientation chosen on $M$ and in fact one can
introduce a different orientation. The numerical factor is included for
convenience of simplifying various relations below. Such a formally defined
volume is complex, in general (for the Lorentzian geometry it is purely
imaginary, see below).

Both relations (\ref{nodeg}) and (\ref{dcompletc}) are explicitly covariant
with respect to the $SL(2,C)$ transformations defined on the objects with
the ``undotted spinor indices'' in accordance with 
\begin{equation}
\iota\longrightarrow\ell\cdot\iota:\quad\quad
(\ell\cdot\iota)^{A}=\ell^{A}_{B}\, \iota^{B},\;
(\ell\cdot\iota)_{B}= \ell^{A}_{B}\, \iota_{A},\label{untra}
\end{equation}
where the complex transformation matrix $\vert \ell^{A}_{B}\vert$ obeys
the only constraint 
$$
\epsilon^{CD}\epsilon_{AB}\,\ell^{A}_{C}\,\ell^{B}_{D}=2,
$$
i.e. is unimodular, $|\ell^{A}_{B}|\in SL(2,C)$.

We shall name the 2-forms $S_{AB}$ whose components obey the above
conditions $S$-{\it forms} for the sake of brevity.
\medskip

{\bf Proposition 1}\ {\it For each set of $S$-forms which 
satisfy the completeness} (\ref{dcompletc}) {\it and the non-degeneracy} 
(\ref{nodeg}) {\it conditions, there exists a basis $\theta_{A\dot{B}}$ 
of the complexified cotangent bundle $\Lambda^1$ such that}
\begin{equation}
S_{AB}=\textstyle{1\over 2}\theta_{A}{}^{\dot K}\wedge\theta_{B\dot{K}}.
\label{SvT}
\end{equation}

\noindent {\it Proof:}\quad Eqs.~(\ref{dcompletc}) include in
particular the equations $S_{00}\wedge S_{00}=S_{11}\wedge S_{11}=0$,
$S_{00}\wedge S_{11}=-\eta$, which imply an existence of the linearly 
independent 1-forms $\theta^j,\;j=1,2,3,4,$ such that $S_{00}=\theta^3
\wedge\theta^1$, $S_{11}=\theta^4\wedge\theta^2$, $\theta^1\wedge\theta^2
\wedge\theta^3\wedge\theta^4=\eta\not=0$. The latter relations are not 
changed under the transformations of the 1-forms,
\begin{equation}
\pmatrix{\theta^3\cr \theta^1\cr}\longrightarrow \ell_{(1)}\cdot
\pmatrix{\theta^3\cr \theta^1\cr},\quad\quad 
\pmatrix{\theta^4\cr \theta^2\cr}\longrightarrow \ell_{(2)}\cdot
\pmatrix{\theta^4\cr \theta^2\cr},\label{rot}
\end{equation}
where $\ell_{(1)},\ell_{(2)}$ are arbitrary complex $2\times2$ unimodular 
matrices. On the account of the further equations (\ref{dcompletc}),
$S_{01}\wedge S_{00}=S_{01}\wedge S_{11}=0$, $S_{01}\wedge S_{01}=
{1\over 2}\eta$, one notices that the 2-form $S_{01}=S_{10}$ admits the 
expansion $S_{01} = a\theta^3\wedge\theta^2 + b\theta^4\wedge\theta^1 +
p\theta^1\wedge\theta^2 + q\theta^3\wedge\theta^4$, 
where the coefficients $a,b,p,q$ satisfy $ab+pq={1\over 4}$. Applying the 
$SL(2,C)$ transformations (\ref{rot}) one can reduce this to $-2S_{01}= 
\theta^1\wedge\theta^2 + \theta^3\wedge\theta^4$. Finally, introducing the 
spinor indexing of $\theta$-tetrad by means of the correspondence 
\begin{equation}
\theta^3\mapsto\theta_{0\dot{0}},\quad\theta^1\mapsto\theta_{0\dot{1}},\quad
\theta^2\mapsto\theta_{1\dot{0}},\quad\theta^4\mapsto -\theta_{1\dot{1}},
\end{equation}
Eq.~(\ref{SvT}) immediately follows.$\Box$
 \medskip

{\bf Corollary 2}\ {\it Given the 1-forms $\theta_{A\dot{B}}$ from the 
proposition 1, a family of 2-forms $S_{\dot{A}\dot{B}}=
S_{\dot{B}\dot{A}}$, defined by}
\begin{equation}
S_{\dot{A}\dot{B}}=\textstyle{1\over 2}\theta^{K}{}_{\dot{A}}\wedge
\theta_{K\dot{B}}\label{dSvT}
\end{equation}
{\it and consisting of three elements $S_{\dot{0}\dot{0}}, 
S_{\dot{0}\dot{1}}=S_{\dot{1}\dot{0}}, S_{\dot{1}\dot{1}}$, obeys the 
``dotted'' version of eqs.}~(\ref{dcompletc}), 
\begin{equation}
S^{\dot{A}\dot{B}}\wedge S_{\dot{C}\dot{D}}=\textstyle{1\over 3}
\delta^{\dot A}_{(\dot B}\,\delta^{\dot C}_{\dot D)}\;
S^{\dot{K}\dot{L}}\wedge S_{\dot{K}\dot{L}},\label{completc}
\end{equation}
{\it and the identities}
\begin{equation}
S_{AB}\wedge S_{\dot{C}\dot{D}}=0, \quad S^{KL}\wedge S_{KL} + 
S^{\dot{K}\dot{L}}\wedge S_{\dot{K}\dot{L}}= 0.\label{SdSid}
\end{equation}

\noindent {\it Proof:} The most straightforward way to prove eqs.
(\ref{completc})-(\ref{SdSid}) is to notice that $S_{AB}$ and 
$S_{\dot{A}\dot{B}}$ together form the basis 
$\theta_{A\dot{B}}\wedge\theta_{C\dot{D}}$ of the complexified 2-forms 
$\Lambda^2$ bundle as follows:
\begin{equation}
\theta_{A\dot{B}}\wedge\theta_{C\dot{D}}=\epsilon_{AC}
S_{\dot{B}\dot{D}} + \epsilon_{\dot{B}\dot{D}}S_{AC}.\label{ttSS}
\end{equation}
The equation (\ref{ttSS}) is directly proved with the help of the standard 
spinor algebra methods.$\Box$

Notice that all the above equations are invariant with respect to another
independent copy of the group $SL(2,C)$ besides (\ref{untra}). The second
$SL(2,C)$ transformation group acts on ``dotted spinor indices'' by the
similar rules
$\iota^{\dot A}\rightarrow\ell^{\dot A}_{\dot B}\, \iota^{\dot B},\;
\iota_{\dot B}\rightarrow \ell^{\dot A}_{\dot B}\, \iota_{\dot A}$,
where $\vert \ell^{\dot A}_{\dot B}\vert\in SL(2,C)$. In general, these
two transformation groups are unrelated to each other.

An immediate consequence of the proposition 1 is the existence of metric on
$M$ which can be defined in an $SL(2,C)$ invariant manner as 
$g=\theta_{A}{}^{\dot{B}}\otimes\theta^{A}{}_{\dot{B}}$. In the local
coordinates $x^{\alpha}$ the complex frame has the components 
$\theta^{A\dot{B}}{}_{\alpha}=\partial_{\alpha}
{\_\!\_\kern -0.2ex\raise 0.4ex\hbox{$\scriptstyle |$}}
\theta^{A\dot{B}}$ and hence the components of the metric are
\begin{equation}
g_{\alpha\beta}=
\theta_{A}{}^{\dot{B}}{}_{\alpha}\theta^{A}{}_{\dot{B}\beta}.\label{met}
\end{equation} From 
the proposition 1 it follows that this symmetric tensor is non-degenerate.

Given the volume 4--form (\ref{eta}) and the corresponding coframe 1-form
$\theta^{A}{}_{\dot{B}}$ (metric $g_{\alpha\beta}$), one can define the 
$\eta$-basis of the exterior algebra and compute the Hodge duals. Let us 
denote $e_{A}{}^{\dot{B}}$ the complex frame (basis of the complexified 
tangent space) dual to the fundamental coframe 1-form, i.e.
\begin{equation}
e_{A}{}^{\dot{B}}{\_\!\_\kern -0.2ex\raise 0.4ex\hbox{$\scriptstyle |$}}
\theta^{C}{}_{\dot{D}}=\delta^{C}_{A}\delta^{\dot B}_{\dot D}.\label{dual}
\end{equation}
\medskip

{\bf Proposition 3}\ {\it The Hodge duals and the $\eta$-basis related to 
the coframe $\theta^{A}{}_{\dot{B}}$ are as follows}:
\begin{eqnarray}
\eta_{A}{}^{\dot{K}}&=&e_{A}{}^{\dot{B}}
{\_\!\_\kern -0.2ex\raise 0.4ex\hbox{$\scriptstyle |$}}\eta =
\ast\theta_{A}{}^{\dot{K}}\nonumber\\ 
&=&\textstyle{2\over 3}S_{AC}\wedge\theta^{C\dot{K}},\label{inv}
\end{eqnarray}
\begin{eqnarray}
\eta_{AB}{}^{\dot{K}\dot{L}}&=&e_{B}{}^{\dot{L}}
{\_\!\_\kern -0.2ex\raise 0.4ex\hbox{$\scriptstyle |$}}\eta_{A}{}^{\dot{K}}
=\ast(\theta_{A}{}^{\dot{K}}\wedge\theta_{B}{}^{\dot{L}})\nonumber\\
&=&\epsilon_{AB}S^{\dot{K}\dot{L}} - \epsilon^{\dot{K}\dot{L}}S_{AB},
\label{du2}
\end{eqnarray}
\begin{eqnarray}
\eta_{ABC}{}^{\dot{K}\dot{L}\dot{M}}&=&e_{C}{}^{\dot{M}}
{\_\!\_\kern -0.2ex\raise 0.4ex\hbox{$\scriptstyle |$}}
\eta_{AB}{}^{\dot{K}\dot{L}}=\ast(\theta_{A}{}^{\dot{K}}\wedge
\theta_{B}{}^{\dot{L}}\wedge\theta_{C}{}^{\dot{M}}) \nonumber \\&=&
\epsilon_{AB}\epsilon^{\dot{M}(\dot{L}}\theta_{C}{}^{\dot{K})} -
\epsilon^{\dot{K}\dot{L}}\epsilon_{C(B}\theta_{A)}{}^{\dot{M}},\label{du3}
\end{eqnarray}
\begin{eqnarray}
\eta_{ABCD}{}^{\dot{K}\dot{L}\dot{M}\dot{N}}&=&e_{D}{}^{\dot{N}}
{\_\!\_\kern -0.2ex\raise 0.4ex\hbox{$\scriptstyle |$}}
\eta_{ABC}{}^{\dot{K}\dot{L}\dot{M}}=\ast(\theta_{A}{}^{\dot{K}}\wedge
\theta_{B}{}^{\dot{L}}\wedge\theta_{C}{}^{\dot{M}}\wedge
\theta_{D}{}^{\dot{N}})\nonumber\\&=&- \epsilon_{AB}\epsilon_{CD}
\epsilon^{\dot{M}(\dot{K}}\epsilon^{\dot{L})\dot{N}}+ 
\epsilon_{D(A}\epsilon_{B)C}\epsilon^{\dot{K}\dot{L}}
\epsilon^{\dot{M}\dot{N}}.\label{du4}
\end{eqnarray}

\noindent {\it Proof:} One should straightforwardly calculate subsequent
interior products. For example, starting from (\ref{eta}) and using
(\ref{SvT}) and (\ref{dual}), one finds
$$
e_{A}{}^{\dot{B}}{\_\!\_\kern -0.2ex\raise 0.4ex\hbox{$\scriptstyle |$}}
\eta = \textstyle{1\over 3}S_{KL}\wedge e_{A}{}^{\dot{B}}
{\_\!\_\kern -0.2ex\raise 0.4ex\hbox{$\scriptstyle |$}} 
(\theta^{K}{}_{\dot{M}}\wedge\theta^{L\dot{M}})=
\textstyle{2\over 3}S_{AL}\wedge\theta^{L\dot{B}}.
$$
This proves (\ref{inv}), the rest relations (\ref{du2})-(\ref{du4}) are 
demonstrated analogously.$\Box$

\medskip

{\bf Corollary 4}\ {\it 
Under the conditions of the proposition 1 the 2-forms $S_{AB}$ and 
$S_{\dot{A}\dot{B}}$ are, respectively, anti-self-dual and self-dual with 
respect to the metric} (\ref{met}) {\it they define},
\begin{equation}
\ast S_{AB}=-S_{AB},\quad\quad \ast S_{\dot{A}\dot{B}}=S_{\dot{A}\dot{B}}.
\label{self}
\end{equation}

\noindent Two contractions of (\ref{du2}) yield (\ref{self}).$\Box$
\medskip

{\bf Corollary 5}\ {\it 
Under the conditions of the proposition 1 the components of the metric}
(\ref{met}) {\it and its determinant are expressed in terms of the rational 
functions of the components of 2-forms $S_{AB}$ via the Urbantke formula}, 
\begin{equation}
g_{\alpha\beta}=\textstyle{2\over 3}\!\ast\!(S_{A}{}^{B}{}_{\alpha}\wedge 
S_{B}{}^{C}\wedge S_{C}{}^{A}{}_{\beta}),\label{urban}
\end{equation}
{\it where} $S_{AB\alpha}=\partial_{\alpha}
{\_\!\_\kern -0.2ex\raise 0.4ex\hbox{$\scriptstyle |$}} S_{AB}$, 
\begin{equation}
3= - \ast\! (S^{KL}\wedge S_{KL}).\label{det}
\end{equation}

\noindent {\it Proof:} Noticing that 
\begin{equation}
S_{AB\alpha}=
\theta_{(A}{}^{\dot{K}}{}_{|\alpha|}\theta_{B)\dot{K}}\label{sa}
\end{equation}
and using (\ref{SvT}), one should apply (\ref{du4}) to the r.h.s. of 
(\ref{urban}). Analogously, making complete pairwise contraction of all 
indices in (\ref{du4}), one finds (\ref{det}).$\Box$

It seems worthwhile to write (\ref{urban}) and (\ref{det}) explicitly in
terms of the $S$--forms components. Recalling that $S_{AB}=S_{AB\mu\nu}
dx^{\mu}\wedge dx^{\nu}$, and hence $S_{AB\alpha}=2S_{AB\alpha\beta}
dx^{\beta}$, one finds for (\ref{urban}) and (\ref{det}), respectively,
\begin{equation}
\sqrt{\det g}g_{\alpha\beta}=-\textstyle{8\over 3}
\varepsilon^{\mu\nu\rho\sigma}S_{A}{}^{B}{}_{\alpha\mu}
S_{B}{}^{C}{}_{\rho\sigma}S_{C}{}^{A}{}_{\nu\beta},
\label{urbanco}
\end{equation}
\begin{equation}
3\sqrt{\det g}= - \varepsilon^{\mu\nu\rho\sigma}S^{KL}{}_{\mu\nu}
S_{KL\rho\sigma}.\label{detco}
\end{equation}
Hence we see that indeed the metric components and its determinant are the
rational functions of components of $S$--forms. One should keep in mind that
the signs on the r.h.s.'s of (\ref{urbanco})-(\ref{detco}) depend on the
orientation chosen on $M$. With another (opposite) choice both signs would
be pluses. A different, purely algebraic mechanism which changes these signs
is provided by what can be called an ``anti-dotting'' operation. In simple
terms, this means that the dotted $S$-forms $S_{\dot{A}\dot{B}}$ replace
the original $S_{AB}$ in all formulas, and vice versa. This reflects the
completely equivalent role of dotted and undotted $S$-forms in determining
the metric structure.

The 3-forms $\eta_{A\dot{B}}$ indeed form the basis of the space of all 
complex 3-forms over $M$, and this is expressed in the identity
\begin{equation}
\theta^{A}{}_{\dot{K}}\wedge\theta^{B}{}_{\dot{L}}\wedge
\theta^{C}{}_{\dot{M}}=\epsilon^{A(B}\eta^{C)}{}_{\dot{K}}
\epsilon_{\dot{L}\dot{M}}-\epsilon_{\dot{K}(\dot{L}}
\eta^{A}{}_{\dot{M})}\epsilon^{BC}.\label{3bas}
\end{equation}
The latter can be proven with the standard spinor indices manipulation 
tricks after noticing that the l.h.s. is antisymmetric under the 
permutation of any pair of indices $\{^{A}_{\dot{K}}\}, \{^{B}_{\dot{L}}\},
\{^{C}_{\dot{M}}\}$.

The forms $\eta_{A\dot{B}}$ are of particular importance, because their
components actually determine the inverse of the metric tensor (\ref{met}).
Indeed, let us first notice that
\begin{equation}
\theta^{A}{}_{\dot{B}}\wedge\eta_{C}{}^{\dot{D}}=
\delta^{A}_{C}\delta^{\dot{D}}_{\dot{B}}\ \eta,\label{tetr}
\end{equation}
which follows when computing the interior product $e_{C}{}^{\dot{D}}
{\_\!\_\kern -0.2ex\raise 0.4ex\hbox{$\scriptstyle |$}}
(\theta^{A}{}_{\dot{B}}\wedge\eta)$ for the zero 5-form inside the 
parentheses. On the other hand, when taking explicitly the coordinate basis 
$e_{\alpha}=\partial_{\alpha}$ and the corresponding coordinate coframe 
$dx^{\alpha}$ one finds analogously $dx^{\alpha}\wedge\eta_{\beta}=
\delta^{\alpha}_{\beta}\eta$ with
\begin{equation}
\eta_{\alpha}=\partial_{\alpha}
{\_\!\_\kern -0.2ex\raise 0.4ex\hbox{$\scriptstyle |$}}\eta\label{coret}
\end{equation}
as the coordinate 3-form of $\eta$-basis. Expanding $\eta_{A}{}^{\dot{B}}$
with respect to the coordinate basis (\ref{coret}),
\begin{equation}
\eta_{A}{}^{\dot{B}}=e_{A}{}^{\dot{B}\alpha}\eta_{\alpha},
\end{equation}
we immediately see that (\ref{tetr}) yields
\begin{equation}
e_{A}{}^{\dot{B}\alpha}\theta^{C}{}_{\dot{D}\alpha}=
\delta^{C}_{A}\delta^{\dot{B}}_{\dot{D}}.\label{con1}
\end{equation} From this one evidently has
\begin{equation}
e_{A}{}^{\dot{B}\alpha}\theta^{A}{}_{\dot{B}\beta}=
\delta^{\alpha}_{\beta},\label{con2}
\end{equation}
and consequently, the components of the inverse metric tensor are given by
\begin{equation}
g^{\alpha\beta}=
e_{A}{}^{\dot{B}}{}^{\alpha}e^{A}{}_{\dot{B}}{}^{\beta}.\label{inmet}
\end{equation}
It is straightforward to prove the identity $g^{\alpha\beta}g_{\beta\gamma}=
\delta^{\alpha}_{\gamma}$ substituting the definitions (\ref{met}) and
(\ref{inmet}) and (\ref{con1})-(\ref{con2}). Contracting (\ref{inmet}) with
the components of the coframe, and (\ref{met}) with that of the dual frame,
one finds the useful relations,
\begin{equation}
g^{\alpha\beta}\theta_{A}{}^{\dot{B}}{}_{\beta}=e_{A}{}^{\dot{B}\alpha},
\quad g_{\alpha\beta}e_{A}{}^{\dot{B}\beta}=
\theta_{A}{}^{\dot{B}}{}_{\alpha}.\label{move}
\end{equation}

It is clear that the components of $\eta_{A}{}^{\dot{B}}$ describe the 
expansion of the coordinate coframe with respect to the fundamental 
$\theta^{A}{}_{\dot{B}}$ coframe, namely
\begin{equation}
dx^{\alpha}=e_{A}{}^{\dot{B}\alpha}\theta^{A}{}_{\dot{B}}.\label{dx}
\end{equation}
[Let us mention also that the frame of tangent space evidently reads 
$e_{A}{}^{\dot{B}}=e_{A}{}^{\dot{B}\alpha}\partial_{\alpha}$, this 
explains the notation for the components $e_{A}{}^{\dot{B}\alpha}$].

It is straightforward to see that the components of the inverse metric
(\ref{inmet}) are, like (\ref{urbanco})-(\ref{detco}), the rational 
functions of the $S$--forms components, namely,
\begin{equation}
\textstyle{3\over 4}(\det g)\ g^{\alpha\beta} = - 
\varepsilon^{\alpha\mu\nu\gamma}\varepsilon^{\beta\rho\sigma\delta}
S_{A}{}^{B}{}_{\mu\nu}S_{B}{}^{C}{}_{\gamma\delta}
S_{C}{}^{A}{}_{\rho\sigma}.\label{inmetco}
\end{equation}
It seems worthwhile to mention that unlike the relatively well known
so-called Urbantke formula (\ref{urbanco}), the inverse metric expression
(\ref{inmetco}) was not reported in the literature, at least to our
knowledge. Both formulas play an important role in the subsequent 
reformulation of the vacuum Einstein equations in terms of the curvature 
components.

It is clear that $g_{\mu\nu}$, $g^{\mu\nu}$ may be regarded as metric 
tensors. In general, they are complex. However, if the equation
\begin{equation}
S_{AB}\wedge\overline{S_{CD}}=0        
\end{equation}
is fulfilled (where the overbar denotes complex conjugation),  
i.e. the subspaces of $\Lambda^2$ spanned by the triads
$S_{AB}$ and $\overline{S_{AB}}$, respectively, are ``wedge orthogonal'', 
then the metric is conformal to real one possessing the Lorentz signature. 
Moreover, if $\Re (S^{KL}\wedge S_{KL})=0$ the conformal factor reduces to 
one of the fourth roots of 1, i.e. $1,-1,i$ or $-i$. Then the obvious 
constant rescaling of $S$-forms leads to the real Lorentzian metric.

We shall not give a proof of that claim here but mention that in the
latter case $(\det g)$ is real negative. Bearing this in mind, we remark
that perhaps it would be better to rewrite all the component formulas, 
namely, (\ref{urbanco}), (\ref{detco}) and (\ref{inmetco}), with the metric 
determinant in the form $\sqrt{\det g}=i\sqrt{-\det g}$. The definition of
the volume 4-form should be then $\eta=\pm i {1\over 3}S^{KL}\wedge S_{KL}$
instead of (\ref{eta}), with the Hodge duals in the proposition 3 
modified correspondingly. Then the imaginary unit would take its usual 
place in the duality equations (\ref{self}). However, the use of the 
complex {\it formally} Euclidean definitions simplifys all the calculations
below, and we thus will not change the original notation as soon as the 
reality conditions are not imposed until the very end.

The 1--forms $S_{AB\alpha}$ prove to be an extremely useful objects in the 
subsequent discussion of relations between $S$-forms and connection and
curvature forms. The following statement summarizes some of the key
properties of these forms.
\medskip

{\bf Proposition 6}\ {\it 
Under the conditions of the proposition 1, the 1-forms $S_{AB\alpha}$ 
satisfy the following identities}
\begin{equation}
S^{AB}{}_{\alpha}\wedge S_{CD}= \delta^{(A}_{(C}\ast\! S^{B)}{}_{D)\alpha}
- \textstyle{1\over 2}\delta^{A}_{(C}\delta^{B}_{D)}\ \eta_{\alpha},
\label{mainA}
\end{equation}
\begin{equation}
S_{CD}{}_{\alpha}\ast\!(dx^{\alpha}\wedge dx^{\beta}\wedge S^{AB}) = -
\ast(dx^{\beta}\wedge S^{(A}{}_{(C}\delta^{B)}_{D)})-\textstyle{1\over 2}
\delta^{A}_{(C}\delta^{B}_{D)}\ dx^{\beta},\label{mainB}
\end{equation}
{\it where, as usually in this paper, the Hodge star dual is determined by 
the metric defined by the $S$--forms (as summarized in} 
(\ref{inv})-(\ref{du4})).

\noindent {\it Proof:} We start directly from the completeness condition 
and apply the interior product with the coordinate basis 
$\partial_{\alpha}$ to (\ref{dcompletc}). This yields
\begin{equation}
S^{AB}{}_{\alpha}\wedge S_{CD} + S^{AB}\wedge S_{CD}{}_{\alpha}= 
- \delta^{A}_{(C}\delta^{B}_{D)}\ \eta_{\alpha},\label{main1}
\end{equation}
where we have used (\ref{eta}) and (\ref{coret}). Now from (\ref{3bas}) 
we find
\begin{equation}
\theta^{A}{}_{\dot{D}}\wedge S_{BC} = 
\delta^{A}_{(B}\eta_{C)\dot{D}},\label{ths}
\end{equation}
and using (\ref{sa}), one straightforwardly computes 
\begin{equation}
S^{AB}{}_{\alpha}\wedge S_{CD} - S^{AB}\wedge S_{CD}{}_{\alpha}= 
\theta^{(A|\dot{K}|}{}_{\alpha}\delta^{B)}_{(C}\wedge\eta_{D)\dot{K}} 
+\theta_{(C}{}^{\dot{K}}{}_{|\alpha|}\delta^{(A}_{D)}\wedge
\eta^{B)}{}_{\dot{K}}.\label{main2}
\end{equation}
Applying Hodge operator to (\ref{sa}), one gets
\begin{equation}
\ast S_{AB\alpha}=
\theta_{(A}{}^{\dot{K}}{}_{|\alpha|}\eta_{B)\dot{K}},\label{sadu}
\end{equation}
and it remains only to rearrange the r.h.s. of (\ref{main2}) with the 
help of (\ref{sadu}), with the final formula
\begin{equation}
S^{AB}{}_{\alpha}\wedge S_{CD} - S^{AB}\wedge S_{CD}{}_{\alpha}= 
2 \delta^{(A}_{(C}\ast\! S^{B)}{}_{D)\alpha}.\label{main3}
\end{equation}
Combining (\ref{main1}) and (\ref{main3}), one proves the first identity 
(\ref{mainA}).

The second identity (\ref{mainB}) is simply the Hodge dual of the first one,
although some efforts are required to demonstrate this explicitly. To begin
with, one easily notices that since $\eta_{\alpha}=\ast dx_{\alpha}$
(hereafter lower coordinate index is understood to be moved with the help
of the metric $g_{\alpha\beta}$), the last terms on the r.h.sides of 
(\ref{mainA}) and (\ref{mainB}) are dual. As for the first terms on the 
r.h.sides, one derives
\begin{equation}
dx_{\beta}\wedge S_{BC}=g_{\alpha\beta}dx^{\alpha}\wedge S_{BC}=
g_{\alpha\beta}e_{A}{}^{\dot{D}\alpha}\theta^{A}{}_{\dot{D}}\wedge 
S_{BC}=\theta_{(B}{}^{\dot{D}}{}_{|\beta|}\eta_{C)\dot{D}},\label{dual1}
\end{equation}
where we used (\ref{dx}), (\ref{move}), and (\ref{ths}). Thus
in view of (\ref{sadu}),
\begin{equation}
\ast (dx_{\beta}\wedge S_{BC}) = - S_{BC\beta},\label{dual2}
\end{equation}
and we have proven that the r.h.s. of (\ref{mainA}) is a Hodge dual of the
r.h.s. of (\ref{mainB}). Now let us prove the duality of the left hand 
sides. We will several times make use of the well known identity 
\begin{equation}
\Phi\wedge\ast\Psi = \Psi\wedge\ast\Phi,\label{prod}
\end{equation}
which holds for any forms $\Phi, \Psi$ of equal degree. As a preliminary
step we notice that
\begin{equation}
dx_{\gamma}\wedge S^{AB}{}_{\beta}\wedge S_{CD} =
dx_{\beta}\wedge S_{CD\gamma}\wedge S^{AB}.\label{step1}
\end{equation}
This is directly seen when multiplying (\ref{mainA}) by the coordinate 
coframe 1-form from the left, and then recalling that $dx^{\gamma}\wedge
\eta_{\beta}=\delta^{\gamma}_{\beta}\eta = dx_{\beta}\wedge\eta^{\gamma}$, 
while using (\ref{prod}),
\begin{eqnarray}
dx_{\gamma}\wedge\ast S_{BD\beta}&=& S_{BD\beta}\wedge\eta_{\gamma}=
2S_{BD\beta\alpha}dx^{\alpha}\wedge\eta_{\gamma}=
2S_{BD\beta\gamma}\eta \label{step2}\\
&=& - dx_{\beta}\wedge\ast S_{BD\gamma}.\nonumber
\end{eqnarray}
And now we can complete the demonstration. By using repeatedly (\ref{prod})
and (\ref{step2}), we find
\begin{eqnarray}
dx_{\gamma}\wedge\ast\big[ S_{CD\alpha}\ast\!(dx^{\alpha}\wedge dx^{\beta}
\wedge S^{AB})\big] &=& \big[ S_{CD\alpha}\ast\!(dx^{\alpha}\wedge 
dx^{\beta}\wedge S^{AB})\big]\wedge\eta_{\gamma}\nonumber\\ &=& 
2S_{CD\alpha\gamma}\ast\!(dx^{\alpha}\wedge dx^{\beta}\wedge S^{AB})
\eta\nonumber\\ &=& 2S_{CD\alpha\gamma}dx^{\alpha}\wedge dx^{\beta}
\wedge S^{AB}\nonumber\\ &=& dx^{\beta}\wedge S_{CD\gamma}\wedge S^{AB}
\nonumber\\ &=& dx_{\gamma}\wedge\big[ S^{AB\beta}\wedge S_{CD}\big],
\label{main4}
\end{eqnarray}
where we have used (\ref{step1}) at the last line. Thus by Cartan's lemma
we see that
$$
S^{AB}{}_{\beta}\wedge S_{CD}=\ast\big[ S_{CD\alpha}\ast\!(dx^{\alpha}
\wedge dx_{\beta}\wedge S^{AB})\big],
$$
and this ends the proof of the second identity (\ref{mainB}).$\Box$
\medskip

{\bf Corollary 7}\ {\it An arbitrary spinor-valued symmetric 1--form 
$T_{AB}=T_{BA}$ and any 1-form $\psi$ satisfy the identities} 
\begin{equation}
\ast(\psi\wedge S_{AB})\wedge S^{CD} = \psi\wedge S^{(C}_{\ \ (A}
\delta^{D)}_{B)} + \textstyle{1\over 2}\ast\!\psi\ \delta^{C}_{(A}
\delta^{D}_{B)},\label{forcon3}
\end{equation}
\begin{equation}
T_{AB}=-\left(\ast\!(dx^{\alpha}\wedge U_{AC})S^{C}{}_{B\alpha} + 
\ast(dx^{\alpha}\wedge U_{BC})S^{C}{}_{A\alpha}\right)
+ \ast U_{AB},\label{forcon}
\end{equation}
\begin{equation}
T_{AB}\wedge S^{CD}= - U^{(C}_{\ \ (A}\delta^{D)}_{B)} + 
\ast (U_{AB})\wedge S^{CD} + \ast (U^{(C}_{\ \ (A})\wedge 
S^{D)}_{\ \ B)} + \ast (U^{K}_{\ \ (A})\delta^{(C}_{B)}\wedge 
S^{D)}_{\ \ K},\label{forcon2}
\end{equation}
{\it with}
$$
U_{AB}=T^{D}{}_{(A}\wedge S_{B)D},
$$
{\it where the $S$--forms satisfy the conditions of the proposition 1, 
and the Hodge dual is defined by the $S$-corresponding metric}.

\noindent {\it Proof:}\ Take the coefficients of the $T$-form in coordinate
basis, $T_{AB\beta}dx^{\beta}$ and compute the contraction of the identity
(\ref{mainB}) with $2\delta^{D}_{(M}\epsilon_{N)(A}\delta^{C}_{K)}
T^{K}{}_{B\beta}$. This proves (\ref{forcon}), while (\ref{forcon2}) 
follows from the latter with the help of (\ref{mainA}). The identity
(\ref{forcon3}) is in fact a different form of (\ref{mainA}) with the
relation (\ref{dual2}) inserted.$\Box$

Immediate consequences of (\ref{forcon3})-(\ref{forcon2}) are 
\begin{equation}
\ast(\psi\wedge S_{AB})\wedge S^{AB} = \textstyle{3\over 2}\ast\!\psi,
\label{forbi2}
\end{equation}
\begin{equation}
T_{AB}\wedge S^{AB} = -\ast(T^{D}{}_{(A}\wedge S_{B)D})\wedge S^{AB},
\label{forbi1}
\end{equation}
which are identically satisfied for any symmetric 1-form $T_{AB}$ and any
1-form $\psi$.
\medskip

{\bf Corollary 8}\ {\it Under the conditions of the proposition 1, the 
following cubic identity is fulfilled} (cf. (\ref{urban})),
\begin{equation}
S^{AB}{}_{\alpha}\wedge S_{CD}\wedge S^{EF}{}_{\beta} = \eta\left(
\textstyle{1\over 2}g_{\alpha\beta}\delta^{(A}_{(C}\epsilon^{B)(E}
\delta^{F)}_{D)} - 2 \delta^{(A}_{(C}S^{B)(E}{}_{|\alpha\beta|}
\delta^{F)}_{D)}\right).\label{bSSS}
\end{equation}

\noindent {\it Proof:} Multiply the identity (\ref{mainA}) by 
$S^{EF}{}_{\beta}$ and then use (\ref{prod}), (\ref{step2}) and 
(\ref{mainA}).$\Box$

\section{Algebraic relations of $S$- and $\Omega$-forms}
\label{ASOm}

Let us introduce some convenient notations. We define the quadratic 
($\hbox{\boldmath $I$}$) and cubic ($\hbox{\boldmath $J$}$) ``totally 
contracted'' operators over 
the objects $Z^{ABCD}$ with four spinor indices which are symmetric with 
respect to the first and last pairs, $Z^{ABCD}=Z^{(AB)(CD)}=Z^{(CD)(AB)}$. 
Specifically, let
\begin{eqnarray}
\hbox{\boldmath $I$}[Z]&=&Z_{KL}{}^{MN}\,Z_{MN}{}^{KL},\nonumber\\
\hbox{\boldmath $J$}[Z]&=&Z_{KL}{}^{MN}\,Z_{MN}{}^{PQ}\,
Z_{PQ}{}^{KL},\nonumber\\
\hbox{\boldmath $J$}_{k}[Z]&=&\hbox{\boldmath $J$}[Z]- 
k\ (\mathop{\bf tr}[Z])^3,\quad
{\rm where}\ \mathop{\bf tr}[Z]=Z_{KL}{}^{KL}.\label{defDel}
\end{eqnarray}

Let us consider, preliminarily,  the problem of the inverting of a 
$3\times 3$ matrix formed by the components of a totally symmetric 4-index 
spinor $C^{ABCD}=C^{(ABCD)}$. (A particular example of such an object is 
provided by the Weyl conformal curvature spinor, see also section 
\ref{GeoS}). This problem arises, in particular, in the pure connection 
formulation of general relativity \cite{cdj3} but it was only briefly 
discussed there.
\medskip

{\bf Proposition 9}\ {\it 
Let the 4-index spinor $Z_{ABCD}$ be symmetric in the first and last index
pairs, $Z_{ABCD}=Z_{(AB)(CD)}=Z_{(CD)(AB)}$, and obey the algebraic 
constraints}
\begin{equation}
Z^{K}{}_{(A\,B)K}=0,\quad \hbox{\boldmath $I$}[Z]=
(\mathop{\bf tr}[Z])^2.\label{Zeq}
\end{equation}
{\it Then}
\begin{equation}
{\widetilde C}_{ABCD}=Z_{AB}{}^{KL}\,Z_{KL}{}^{CD} -
Z_{ABCD}\,Z_{KL}{}^{KL}\label{CZ}
\end{equation}
{\it is totally symmetric and obeys the equations}
\begin{eqnarray}
Z_{AB}{}^{KL}\,{\widetilde C}_{KL}{}^{CD}&=& 
\textstyle{1\over 3}\hbox{\boldmath $J$}_{1}[Z]
\delta^C_{(A}\,\delta^D_{B)},\label{ZCort}\\
\textstyle{1\over 3}\hbox{\boldmath $J$}_{1}[Z]\ Z_{AB}{}^{CD}&=&
{\widetilde C}_{AB}{}^{KL}\,{\widetilde C}_{KL}{}^{CD} -
\textstyle{1\over 2}\hbox{\boldmath $I$}[{\widetilde C}]\ \delta^C_{(A}\,
\delta^D_{B)}.\label{CCZ}
\end{eqnarray}

\noindent{\it Proof:}\quad Regarding the index pairs $\scriptstyle AB$ and 
$\scriptstyle CD$ as multi-indices, $Z_{AB}{}^{CD}$ may be interpreted as 
a $3\times3$ matrix $[Z]$. As such, it annihilates its own characteristic
polynomial which equals
$$
[Z]^3-(\mathop{\bf tr}[Z])[Z]^2-\textstyle{1\over 2}\left\{\mathop{\bf tr}
([Z]^2) - (\mathop{\bf tr}[Z])^2\right\}[Z] 
$$
$$
-\textstyle{1\over 6}\left\{(\mathop{\bf tr}[Z])^3 - 3\mathop{\bf tr}[Z]
\mathop{\bf tr}([Z]^2) + 2\mathop{\bf tr}([Z]^3)\right\}[I].
$$
In our case $\mathop{\bf tr}([Z]^2)=\hbox{\boldmath $I$}[Z]$, 
$\mathop{\bf tr}([Z]^3)=\hbox{\boldmath $J$}[Z]$, the unit matrix 
$[I]=|\delta^{A}_{(C}\delta^{B}_{D)}|$. Using the second of eqs. 
(\ref{Zeq}), one finds
$$
[Z]\ \left\{[Z]^2 - [Z]\mathop{\bf tr}[Z]\right\}=
\textstyle{1\over 3}[I]\left\{\hbox{\boldmath $J$}[Z]-
(\mathop{\bf tr}[Z])^3)\right\},
$$
and (\ref{ZCort}) immediately follows.

Next, eqs. (\ref{CZ}), (\ref{ZCort}) yield
$$
{\widetilde C}_{AB}{}^{KL}\,{\widetilde C}_{KL}{}^{CD}= 
\textstyle{1\over 3}\hbox{\boldmath $J$}_{1}[Z](Z_{AB}{}^{CD} - 
Z_{KL}{}^{KL} \delta^C_{(A}\,\delta^D_{B)})
$$
and hence $\hbox{\boldmath $I$}[{\widetilde C}]= - \textstyle{2\over 3}
\mathop{\bf tr}[Z]\hbox{\boldmath $J$}_{1}[Z]$, that proves (\ref{CCZ}).

Finally, the first of eqs.~(\ref{Zeq}) entails the total symmetry of
${\widetilde C}_{ABCD}$: it is easy to show that any contraction of the 
r.h.s. of (\ref{CZ}) vanishes.$\Box$

Now let us assume $Z_{ABCD}$ to be not an arbitrary but constructed from 
the components of a set of 2-forms ${\Omega}_{AB}={\Omega}_{AB\mu\nu}
dx^{\mu}\wedge dx^{\nu}$ obeying the index symmetries of $S$-forms:
${\Omega}_{AB\mu\nu}={\Omega}_{(AB)[\mu\nu]}$. Specifically, let
\begin{equation}
Z_{AB}{}^{CD} ={\Omega}_{AB}{}^{CD} - \textstyle{1\over 2}
\Omega_{KL}{}^{KL}\ \delta^{C}_{(A}\,\delta^{D}_{B)},\label{ZOm}
\end{equation}
where we denoted 
\begin{equation}
{\Omega}_{AB}{}^{CD}=\varepsilon^{\alpha\beta\mu\nu}\,
{\Omega}_{AB\alpha\beta}{\Omega}^{CD}{}_{\mu\nu}.
\end{equation}
Notice that, by construction, $\Omega^{K}{}_{(AB)K}=0$. 

Further, we introduce another set of two-forms,
\begin{equation}
{\widetilde S}_{AB}=Z_{AB}{}^{KL}\,{\Omega}_{KL} = \Omega_{AB}{}^{KL}\,
{\Omega}_{KL} - \textstyle{1\over 2}\Omega_{KL}{}^{KL}\,\Omega_{AB}
\label{StOm}
\end{equation}
which possess the same index symmetries as $\Omega$- and $S$-forms and are 
the homogeneous cubic polynomials in the components ${\Omega}_{AB\mu\nu}$. 
\medskip

{\bf Proposition 10}\ {\it 
If the components ${\Omega}_{AB\mu\nu}$  obey the constraints}
\begin{equation}
2{\Omega}_{AB}{}^{CD}{\Omega}_{CD}{}^{AB}=({\Omega}_{KL}{}^{KL})^2,
\label{qu}
\end{equation}
\begin{equation}
\hbox{\boldmath $J$}_{1\over 4}[\Omega]={\Omega}_{AB}{}^{CD}
{\Omega}_{CD}{}^{PQ}{\Omega}_{PQ}{}^{AB} - 
\textstyle{1\over 4}({\Omega}_{KL}{}^{KL})^3\neq 0, \label{Omids}
\end{equation}
{\it then the 2-forms ${\widetilde S}_{AB}$ satisfy the conditions of the
proposition 1, namely}
\begin{equation}
\widetilde{S}^{AB}\wedge\widetilde{S}_{CD}= \textstyle{1\over 3}
\delta^{A}_{(C}\,\delta^{B}_{D)}\;\widetilde{S}^{KL}\wedge
\widetilde{S}_{KL},\quad \widetilde{S}^{KL}\wedge
\widetilde{S}_{KL}\neq 0.\label{cond1}
\end{equation}

\noindent{\it Proof:}\quad Equation (\ref{qu}) guarantees the satisfaction
of the conditions (\ref{Zeq}) of the proposition 9 for (\ref{ZOm}).
The corresponding totally symmetric spinor reads
\begin{eqnarray}
\widetilde{C}_{CD}{}^{KL}&=&\Omega_{CD}{}^{MN}\Omega_{MN}{}^{KL} -
\textstyle{1\over 2}\Omega_{CD}{}^{KL}\Omega_{PQ}{}^{PQ}\nonumber \\
&=& Z_{CD}{}^{MN}\Omega_{MN}{}^{KL}\label{COm}
\end{eqnarray}
In view of (\ref{ZCort}) and (\ref{COm}) we find
\begin{equation}
\widetilde{S}^{AB}\wedge\widetilde{S}_{CD}= 
\textstyle{1\over 3}\delta^{A}_{(C}\,\delta^{B}_{D)}\
\hbox{\boldmath $J$}_{1}[Z]dx^1\wedge dx^2\wedge dx^3\wedge dx^4 
\end{equation}

It is easy to see that $\mathop{\bf tr}[Z]=-{1\over 2}\Omega_{KL}{}^{KL}$,
$\hbox{\boldmath $I$}[Z]={1\over 4}({\Omega}_{KL}{}^{KL})^2$, and 
$\hbox{\boldmath $J$}_{1}[Z]=\hbox{\boldmath $J$}_{1\over 4}[\Omega]$, hence
\begin{equation}
\widetilde{S}^{KL}\wedge\widetilde{S}_{KL}= \hbox{\boldmath $J$}_{1\over 4}
[\Omega]dx^1\wedge dx^2\wedge dx^3\wedge dx^4 \neq 0.\label{st2}
\end{equation}
This ends the proof.$\Box$

These results solve the problem of inverting the relation (\ref{StOm}). 
The inverse reads
\begin{equation}
\textstyle{1\over 3}\hbox{\boldmath $J$}_{1\over 4}[\Omega]{\Omega}_{AB}=
{\widetilde C}_{AB}{}^{KL}\,\widetilde{S}_{KL}.
\end{equation}
For further applications we now introduce new objects
\begin{equation}
S_{AB}=e^{\phi}\,\widetilde{S}_{AB},\quad {C}_{ABCD}=3 e^{-\phi}
(\hbox{\boldmath $J$}_{1\over 4}[\Omega])^{-1}
{\widetilde C}_{ABCD} \label{CtC}
\end{equation}
with some yet unspecified scalar $\phi$. Obviously,
\begin{equation}
{\Omega}_{AB}= C_{AB}{}^{KL}\,S_{KL}.   \label{VEq}
\end{equation}
Notice that the 2-forms $S_{AB}$ evidently also satisfy the conditions
of the proposition 1 with
\begin{equation}
S^{KL}\wedge S_{KL}= e^{2\phi}\hbox{\boldmath $J$}_{1\over 4}[\Omega]
dx^1\wedge dx^2\wedge dx^3\wedge dx^4 .\label{ISex}
\end{equation}

Let us give also the following useful higher order identities:
\begin{eqnarray}
{\Omega}_{AB}{}^{KL}{\widetilde C}_{KL}{}^{CD}&=&\textstyle{1\over 2}
\mathop{\bf tr}[{\Omega}]\,{\widetilde C}_{AB}{}^{CD}+\textstyle{1\over 3}
\hbox{\boldmath $J$}_{1\over 4}[\Omega]\delta^{C}_{(A}\,
\delta^{D}_{B)},\label{1c}\\
\Omega_{A}{}^{KL}{}_{B}\widetilde{C}_{KL}{}^{CD}&=&
\textstyle{1\over 3}\hbox{\boldmath $J$}_{1\over 4}[\Omega]
\delta^{C}_{(A}\,\delta^{D}_{B)},\label{4c}\\
{\widetilde C}_{AB}{}^{KL}{\widetilde C}_{KL}{}^{CD}&=&\textstyle{1\over 3}
\hbox{\boldmath $J$}_{1\over 4}[\Omega]{\Omega}_{AB}{}^{CD},\label{2c}
\end{eqnarray}
\begin{equation}
{\widetilde C}_{AB}{}^{KL}{\widetilde C}_{KL}{}^{MN}
{\widetilde C}_{MN}{}^{CD}=\textstyle{1\over 3}
\hbox{\boldmath $J$}_{1\over 4}[\Omega]\left(\textstyle{1\over 2}
\mathop{\bf tr}[{\Omega}]\,{\widetilde C}_{AB}{}^{CD}+
\textstyle{1\over 3}\hbox{\boldmath $J$}_{1\over 4}[\Omega]
\delta^{C}_{(A}\,\delta^{D}_{B)}\right).\label{3c}
\end{equation} From these one finds
\begin{equation}
\hbox{\boldmath $I$}[{\widetilde C}]=\textstyle{1\over 3}
\hbox{\boldmath $J$}_{1\over 4}[\Omega]\mathop{\bf tr}[{\Omega}],\quad
\hbox{\boldmath $J$}[{\widetilde C}]=\textstyle{1\over 3}
(\hbox{\boldmath $J$}_{1\over 4}[\Omega])^2,\quad
\hbox{\boldmath $J$}[C]=9 e^{-3\phi}
(\hbox{\boldmath $J$}_{1\over 4}[\Omega])^{-1}.\label{JCex}
\end{equation}

\section{Lorentzian geometry in terms of $S$-forms:
basic equations} \label{GeoS}

Here we briefly outline the method of describing the 4-dimensional 
Lorentzian geometry in terms of special sets of $S$-forms. Our 
consideration is everywhere local.

As it was shown in the previous sections, the metrical properties of the
Lorentzian 4-dimensional space-time can be exhaustively described by the 
triad of complex spinor-valued 2-forms $S_{AB}=S_{BA}=S_{AB\mu\nu}dx^{\mu}
\wedge dx^{\nu}$ which obey
\begin{itemize}
\item[(i)] the non-degeneracy condition (\ref{nodeg})
\item[(ii)] the completeness condition (\ref{dcompletc})
and
\item[(iii)] the reality conditions
\begin{equation}
S_{AB}\wedge\overline{S_{CD}}=0,\quad\Re(S^{KL}\wedge S_{KL})=0.\label{real}
\end{equation}
\end{itemize}
The very metric can be restored from the $S$-forms by means of the Urbantke
equations (\ref{urbanco}), (\ref{detco}), (\ref{inmetco}).
However such basic characteristics of the geometry as the connection and
the curvature do not require the immediate use of the metric tensors
and can be completely described in terms of the $S$-forms alone. In 
particular, the symmetric (torsion-free) metric-compatible connection is 
described by the complex-valued 1-forms $\Gamma_{AB}=\Gamma_{BA}$ obeying 
the first structure equations
\begin{equation}
d\,S_{AB} + 2{\Gamma}^{K}{}_{(A}\wedge S_{B)K}=0.\label{Feq}
\end{equation}
The complex-valued 2-forms ${\Omega}_{AB}={\Omega}_{BA}$ of the the 
curvature associated with connection ${\Gamma}_{AB}$ are expressed in terms
of the latter as follows (the second structure equations):
\begin{equation}
{\Omega}^{A}{}_{B}= d\,{\Gamma}^{A}{}_{B}+
{\Gamma}^{K}{}_{B}\wedge {\Gamma}^{A}{}_{K}.\label{Seq}
\end{equation}
Curvature and connection forms satisfy also the Bianchi identities
\begin{equation}
d\,{\Omega}_{AB} + 2{\Gamma}^{K}{}_{(A}\wedge{\Omega}_{B)K}=0.\label{Bid}
\end{equation}
As it is well known, the {\it vacuum Einstein equations} are equivalent to 
the existence of the expansion (cf., e.g., \cite{p1,cdj1,cdj3})
\begin{equation}
{\Omega}_{AB}=Y_{AB}{}^{KL}\,S_{KL} \label{OYex}
\end{equation}
for some coefficients $Y_{ABKL}=Y_{(AB)(KL)}$. The latter can be written
\begin{equation}
Y_{AB}{}^{CD}=\textstyle{1\over 2}{\Psi}_{AB}{}^{CD} -
\textstyle{1\over 12}\delta^{C}_{(A}\delta^{D}_{B)}\,R\;
\end{equation}
in terms of the totally symmetric Weyl spinor (spinor of conformal 
curvature) ${\Psi}_{ABCD}={\Psi}_{(ABCD)}(=2C_{ABCD})$ and the scalar 
curvature $R$ which is constant by virtue of the vacuum version
of Bianchi identities. It follows that $Y^{K}{}_{(A\,B)K}=0$. We shall 
assume $R=0$ below, therefore  restricting our consideration to the case 
of the vanishing cosmological term. Then the relation~(\ref{VEq}), more 
restrictive than (\ref{OYex}), takes place.

Assuming that $\Gamma_{AB}$ and $\Omega_{AB}$ are arbitrary symmetric 
1- and 2-forms (i.e., they do not necessarily satisfy the eqs. (\ref{Seq}) 
and (\ref{Bid})), let us introduce the following differential operators 
\begin{eqnarray}
{{\raisebox{-0.1ex}{$I$}\!\!I}}^{A}{}_{B}&=&
{{\raisebox{-0.1ex}{$I$}\!\!I}}^{A}{}_{B\,\alpha\beta}dx^{\alpha}\wedge 
d\,x^{\beta}=d\,{\Gamma}^{A}{}_{B}+{\Gamma}^{K}{}_{B}\wedge{\Gamma}^{A}
{}_{K}-{\Omega}^{A}{}_{B},\label{IIdef}\\
{{\cal B}}_{AB}&=&{{\cal B}}_{AB\alpha\beta\gamma}\,dx^{\alpha}\wedge 
dx^{\beta}\wedge d\,x^{\gamma}= d\,{\Omega}_{AB} + 
2{\Gamma}^{K}{}_{(A}\wedge{\Omega}_{B)K}.\label{Bidef}
\end{eqnarray}
As it is easily seen, the following identities
\begin{eqnarray}
&&d\,{{\cal B}}_{AB}+2{\Gamma}^{K}{}_{(A}\wedge {{\cal B}}_{B)K}\equiv
2{{\raisebox{-0.1ex}{$I$}\!\!I}}^{K}{}_{(A}\wedge{\Omega}_{B)K},
\label{BiII}\\&&d\,{{\raisebox{-0.1ex}{$I$}\!\!I}}^{A}{}_{B}+
{\Gamma}^{K}{}_{B}\wedge {{\raisebox{-0.1ex}{$I$}\!\!I}}^{A}{}_{K}-
{\Gamma}^{A}{}_{L}\wedge {{\raisebox{-0.1ex}{$I$}\!\!I}}^{L}{}_{B}
\equiv - {{\cal B}}^{A}{}_{B}\label{IIBi}
\end{eqnarray}
hold true for {\it any} ${\Gamma}^{A}{}_{B}$ and ${\Omega}_{AB}$.

Further, assuming the fulfillment of the first structure equation
(\ref{Feq}), one obtains
\begin{equation}
{{\raisebox{-0.1ex}{$I$}\!\!I}}^{K}{}_{(A}\wedge S_{B)K}\equiv 0,
\label{IIS}
\end{equation}
{\it provided} the equation ${\Omega}^{K}{}_{(A}\wedge S_{B)K}=0$
is fulfilled (that holds true in particular in the case of the 
relation~(\ref{OYex})).

\section{Closed form of vacuum Einstein equations}
\label{Eeq}

At first, it would be convenient to define two cubic $\hbox{\boldmath $G$}$ 
operators in a class of the symmetric spinor-valued 2-forms which include, 
in particular, the objects described above as the $S$-forms (cf. eqs. 
(\ref{urbanco}), (\ref{inmetco})):
\begin{eqnarray}
\hbox{\boldmath $G$}[S]_{\alpha\beta}&=&\varepsilon^{\mu\nu\rho\sigma}
S_{A}{}^{B}{}_{\alpha\mu}S_{B}{}^{C}{}_{\rho\sigma}S_{C}{}^{A}
{}_{\nu\beta},\label{bGdef}\\ \hbox{\boldmath $G$}[S]^{\alpha\beta}&=&
\varepsilon^{\alpha\mu\nu\gamma}\varepsilon^{\beta\rho\sigma\delta}
S_{A}{}^{B}{}_{\mu\nu}S_{B}{}^{C}{}_{\gamma\delta}S_{C}{}^{A}
{}_{\rho\sigma}.\label{tGdef}
\end{eqnarray}
Notice that the eqs.~(\ref{bGdef})-(\ref{tGdef}) specify the tensor 
densities with respect to coordinate transformations.

In accordance with the eqs.~(\ref{urban}), (\ref{det}), components of the 
metric tensor can be directly expressed via the corresponding $S$-forms.
On the other hand if the vacuum Einstein equations are fulfilled then the
$S$-forms and the curvature $\Omega$-forms are closely connected by means
of the simple equation~(\ref{VEq}). It may be supposed therefore that the 
metric can be expressed algebraically in terms of the curvature forms as a
rational function of components of latter. Such an algebraic problem was 
investigated in \cite{ihrig,hall,que}, proving this conjecture to be true 
in a wide class of curvature structures. In our approach this is true
provided the {\it generic condition} $\hbox{\boldmath $J$}_{1\over 4}
[\Omega]\neq 0$ is fulfilled, which is equivalent to the non-vanishing of 
the cubic invariant $\hbox{\boldmath $J$}[\Psi]$ of the undotted Weyl 
spinor. However, the algebraic relations alone do not completely suffice 
to determine the metric from the curvature, and a differential equation 
is to be solved to obtain a scalar scaling factor. 

To demonstrate how the curvature is connected with the metric in the vacuum 
case let us calculate the cubic densities $\hbox{\boldmath $G$}
[{\Omega}]_{\mu\nu}$ and $\hbox{\boldmath $G$}[{\Omega}]^{\mu\nu}$. They 
can be easily found using eqs. (\ref{bSSS}) and (\ref{3c}). In the case 
of totally symmetric $Y^{ABCD}=C^{ABCD}$, eq.~(\ref{bSSS}) implies
\begin{equation}
\hbox{\boldmath $G$}[{\Omega}]_{\mu\nu} = \textstyle{1\over 3}
\hbox{\boldmath $J$}[C]\hbox{\boldmath $G$}[S]_{\mu\nu}, 
\quad\hbox{\boldmath $G$}[{\Omega}]^{\mu\nu}=\textstyle{1\over 3}
\hbox{\boldmath $J$}[C]\hbox{\boldmath $G$}[S]^{\mu\nu}.
\end{equation}
At the same time, proposition 10 tells us that the 2-forms 
$\widetilde{S}_{AB}$, which components are directly constructed from the
curvature (\ref{StOm}), also define a metric on $M$. We will denote this
auxiliary metric $\widetilde{g}_{\alpha\beta}$, and the corresponding Hodge 
operator will be also denoted by the tilde, $\widetilde{\ast}$. 

In accordance with eqs. (\ref{st2}), (\ref{JCex}), (\ref{ISex}), one finds
\begin{eqnarray}
g_{\mu\nu} &=& e^{\phi}\widetilde{g}_{\mu\nu},\quad
g^{\mu\nu} = e^{-\phi}\widetilde{g}^{\mu\nu},\label{gtg}\\
\widetilde{g}_{\mu\nu} &=& \textstyle{8\over 3}\hbox{\boldmath $G$}
[{\Omega}]_{\mu\nu},\quad\widetilde{g}^{\mu\nu} = 
-4(\hbox{\boldmath $J$}_{1\over 4}[\Omega])^{-1}
\hbox{\boldmath $G$}[{\Omega}]^{\mu\nu}.\label{gphiOm}
\end{eqnarray}

As we see, the $S$-forms (and the space-time metric $g$) cannot be 
completely determined in a purely algebraic way in terms of the curvature
forms from eq.~(\ref{VEq}) because both $C_{ABKL}$ and $S_{KL}$ are unknown. 
The degree of the corresponding ambiguity is however expressed by a single 
scalar function (generally complex) $\phi$, which is reflected in the fact 
that the eqs.~(\ref{gtg}) involve the yet unspecified factors $e^{\pm\phi}$.
We shall see that $\phi$ can be fixed but differential equations rather 
than algebraic ones have to be drawn here. Specifically, these additional 
equations are the consequence of the of the Bianchi equations.

A possible way of solving the problem of finding the local geometry of a 
generic vacuum space-time from its curvature is described as follows.

Assuming the  fulfillment of conditions (i)-(iii) of the section \ref{GeoS} we
are precisely in a position of the proposition 1 (see section \ref{BL})
and may exploit eqs.~(\ref{mainA}), (\ref{mainB}) (as well as further 
algebraic relations given in sections \ref{BL},\ref{ASOm}). In particular,
the straightforward application of the identity (\ref{forcon}), with an
arbitrary 1-form $T_{AB}$ replaced by the connection, allows to show that 
the only solution of the eq.~(\ref{Feq}) with respect to the connection 
forms $\Gamma_{AB}$ is described by the formula (see earlier discussion in
\cite{ter}):
\begin{equation}
{\Gamma}_{AB}=
-\ast\!(dx^{\alpha}\wedge dS^{K}{}_{(A}) S_{B)K\alpha} -
\textstyle{1\over 2}\ast\! dS_{AB}.\label{explGamma}
\end{equation}

Using (\ref{forcon})-(\ref{forcon3}) and (\ref{1c})-(\ref{JCex}), (after
some lengthy algebra) one finds 
$$
2C^{ABCD}\ast\!(\Gamma^{K}{}_{(A}\wedge\Omega_{B)K})\wedge\Omega_{CD} =
\textstyle{1\over 3}{\hbox{\boldmath $J$}}[C]\Gamma_{AB}\wedge S^{AB},
$$
and hence from (\ref{forbi1}) and the structure equations (\ref{Feq}) 
we get an immediate consequence of the Bianchi identities (\ref{Bid}),
$$
2C^{ABCD}\ast\!(d\Omega_{AB})\wedge\Omega_{CD} 
+\;{\textstyle{1\over 3}}{\hbox{\boldmath $J$}}[C]\ast\! (dS_{AB})
\wedge S^{AB} = 0.
$$
Substituting $C_{ABCD}$ and $S_{AB}$ expressions provided by the eqs.
(\ref{CtC}), and using (\ref{JCex}), one reduces the last equation to
\begin{equation}
d\phi = \Phi[\Omega],\label{phiEq}
\end{equation}
where
\begin{equation}
\Phi[\Omega]:=\textstyle{2\over 3}\;\widetilde{\ast}\!\left(
2\widetilde{C}^{ABCD}\widetilde{\ast}(d\Omega_{AB})\wedge\Omega_{CD}
+ \widetilde{\ast}(d\widetilde{S}_{AB})\wedge\widetilde{S}^{AB}\right).
\label{phidef}
\end{equation}
Here one should keep in mind that ${\widetilde C}_{ABCD}$ and 
${\widetilde S}_{AB}$ are determined by the curvature according to the
eqs. (\ref{COm}), (\ref{StOm}), and all the Hodge duals $\widetilde{\ast}$
are also defined by the curvature with the help of the auxiliary metric 
$\widetilde{g}$ which is explicitly constructed as the rational function of 
the curvature components in accordance with (\ref{gphiOm}).

A remarkable feature of eq.~(\ref{phiEq}) is thus that besides $d\phi$ the 
only functions involved in it are the $\Omega$-forms components. This
differential equation fixes the conformal factor $\phi$. 

We can now finalize the work, rewriting explicitly all the objects and
relations in terms of the curvature. Using the equations (\ref{CtC}), 
(\ref{phiEq}) and (\ref{forbi2}), one finds the closed expression of the 
connection forms in terms of the curvature components and their 
derivatives,
\begin{equation}
{\Gamma}_{AB}= \textstyle{1\over 2}\widetilde{\ast}(\Phi[\Omega]
\wedge\widetilde{S}_{AB}) - \widetilde{\ast}(dx^{\alpha}\wedge 
d\widetilde{S}^{K}{}_{(A})\widetilde{S}_{B)K\alpha} - 
\textstyle{1\over 2}\widetilde{\ast}d\widetilde{S}_{AB}.\label{GammaphiOm}
\end{equation}

A similar transformation (in which (\ref{forbi2}) plays a central role)
re-casts vacuum Bianchi equations (\ref{Bid}) to the form expressed in 
terms of the curvature components alone,
\begin{eqnarray}
{\cal B}[\Omega]_{AB}&\equiv& d\Omega_{AB} + \textstyle{1\over 2}
\Phi[\Omega]\wedge\Omega_{AB} - 
\widetilde{\ast}(d\widetilde{S}^{K}{}_{(A})\wedge\Omega_{B)K}
\nonumber\\&+& {\textstyle{3\over 2}}
{1\over \hbox{\boldmath $J$}_{1\over 4}[\Omega]}\left(
\widetilde{C}_{AB}{}^{KL}d\widetilde{S}_{KL} - 
2\ \widetilde{\ast}(d\widetilde{S}^{KL})\wedge
\widetilde{S}^{M}{}_{(A}\widetilde{C}_{B)KLM}\right)=0.\label{exBia}
\end{eqnarray}

The above facts are summarized in the form of a theorem.
\medskip

{\bf Theorem 11}\ {\it 
In case of non-zero cubic invariant of the conformal curvature
$\hbox{\boldmath $J$}[{\Psi}]={\Psi}^{AB}{}_{CD}{\Psi}^{CD}{}_{EF}
{\Psi}^{EF}{}_{AB}$ the vacuum Einstein equations (with zero cosmological 
term) can be presented in the closed form in terms of the components 
${\Omega}_{AB\mu\nu}$ of the curvature 2-forms ${\Omega}_{AB}=
{\Omega}_{AB\mu\nu}\:dx^{\mu}\wedge dx^{\nu}$. The complete set of 
equations is separated into the following families}:
 \begin{itemize}
\item[(A)]
         {\em algebraic constraints\/}
$$
2{\Omega}^{AB}{}_{CD}\,{\Omega}^{CD}{}_{AB}=({\Omega}^{AB}{}_{AB})^2,
\quad \hbox{\boldmath $J$}_{1\over 4}[\Omega]\neq 0,
$$
{\it where $\Omega_{AB\,CD}=\varepsilon^{\alpha\beta\mu\nu}\,
\Omega_{AB\alpha\beta}\,{\Omega}_{CD\mu\nu},$ and 
$\hbox{\boldmath $J$}_{1\over4}[\ ]$ is defined by} (\ref{defDel});

\item[(B)]
        {\it second order scaling equation}
$$
d(\Phi[\Omega])=0,
$$
{\it where $\Phi[\Omega]$ is defined by} (\ref{phidef}) {\it with 
${\widetilde C}_{ABCD}$ defined by} (\ref{COm}) , {\it and
${\widetilde S}_{AB}$ defined by} (\ref{StOm});

\item[(C)]
    {\it  first order Bianchi equations} (\ref{exBia}) {\it with the Hodge 
duals $\widetilde{\ast}$ defined by the auxiliary metric} (\ref{gphiOm}) 
{\it where $\hbox{\boldmath $G$}[\ ]_{\mu\nu},\hbox{\boldmath $G$}
[\ ]^{\mu\nu}$ are defined by} (\ref{tGdef}), (\ref{bGdef});

\item[(D)]
      {\it second order structure equations}
$$
{\raisebox{-0.1ex}{$I$}\!\!I}[\Omega]^{A}{}_{B}\equiv
d\,{\Gamma[\Omega]}^{A}{}_{B}+
{\Gamma[\Omega]}^{K}{}_{B}\wedge{\Gamma[\Omega]}^{A}{}_{K}
-{\Omega}^{A}{}_{B}=0,
$$
{\it where ${\Gamma[\Omega]}_{AB}$ is defined by} (\ref{GammaphiOm});

\item[(E)]
{\it non-holomorphic reality conditions}
$$
{\Omega}_{AB}\wedge\overline{{\Omega}_{CD}}=0,\quad
\Im{\left(2\Phi[\Omega]+(\hbox{\boldmath $J$}_{1\over 4}[\Omega])^{-1}
d\hbox{\boldmath $J$}_{1\over 4}[\Omega]\right)}=0.
$$
{\it With the item (E) dropped out, the conditions (A)--(D) discriminate 
a complex vacuum solution of Einstein equations}.
\end{itemize}

It can be seen that all the equations mentioned in theorem are
invariant with respect to the two transformation groups:
the group of general coordinate transformations and the group $SL(2,C)$
of spinorial transformations (essentially, its 2-fold covering group 
$SO(3,C)$ which is isomorphic to the special orthochronous Lorentz group).

\noindent{\it Proof\/} of theorem:\quad
The scaling equation (B) implies a local existence of the scalar $\phi$
such that $d\,\phi=\Phi[\Omega]$, determining it up to an arbitrary complex 
constant. By virtue of the algebraic constraints, the eqs.~(\ref{ZOm}),
(\ref{StOm}), (\ref{CtC}) determine a family of 2-forms $S_{AB}=S_{AB\mu\nu}
\,d\,x^{\mu}\wedge d\,x^{\nu}$ which in accordance with the proposition
10 satisfy the conditions of proposition 1, i.e. are the $S$-forms. 
Moreover it can be shown that equations (C),(D) of theorem are nothing else 
but the eqs.~(\ref{Bid}), (\ref{Seq}), respectively, for the connection 
form specified by (\ref{explGamma}), thus automatically obeying (\ref{Feq}).

Concerning the basic field equations listed in the section \ref{GeoS} it
remains to discuss the reality condition
$S_{AB}\wedge\overline{S_{CD}}=0,\ \Re( S_{KL}\wedge S^{KL})=0.$
The first of them is equivalent to the first equation of reality condition 
(E) of theorem since $S_{AB}$ and ${\Omega}_{AB}$ span the same subspaces 
of the complex valued 2-forms space. Further, the second of equations (E) 
is integrated by virtue of the scaling equation and then exponentiated to
$$
\Im{\left(e^{i(\mbox{\scriptsize real constant})}\cdot e^{2\phi}
\hbox{\boldmath $J$}_{1\over 4}[\Omega] \right)}=0.
$$
Notice now that all the equations of theorem are invariant with respect to 
the shift $\phi\rightarrow(\phi\;+$ complex constant) that allows to re-cast 
the above equation to $\Im{(i\,e^{2\phi}\hbox{\boldmath $J$}_{1\over 4}
[\Omega])}\equiv\Re{(e^{2\phi}\hbox{\boldmath $J$}_{1\over 4}[\Omega])}=0$ 
which in its turn coincides with the second reality condition (\ref{real}).
Theorem is therefore proven. $\Box$

It is worthwhile to note that the theorem does not claim a solution of the
equations listed above immediately yields a real metric (by means of the
eqs.~(\ref{gphiOm})). Indeed, it can be seen from the proof that such a
metric may be only {\it conformal} to a real Lorentzian one but
the conformal factor is necessarily a constant (all the equations of the
theorem are invariant with respect to the constant conformal rescaling). 
Then a certain complex shift of the scalar $\phi$ has to be applied to 
provide a real metric. This ambiguity seems however to be unessential and 
so the equations are equivalent to the original vacuum Einstein equations.

\section{Consistency conditions}

The equations listed in theorem 11 are not totally independent but 
manifest some differential and algebraic relations. The latter represent
the adapted form of the general identities (\ref{BiII})-(\ref{IIS}) in 
fact. It is useful to give an explicit form of such a {\it consistency
conditions} of the vacuum Einstein equations which are important in the
construction of the evolution system and for the counting of the
number of degrees of freedom of the field.

It has been mentioned that (\ref{BiII}) and (\ref{IIBi}) hold true for 
arbitrary ${\Gamma}_{AB}$ and ${\Omega}_{AB}$. In particular, the 
connection in terms of the curvature ${\Gamma[\Omega]}_{AB}$, given by 
(\ref{GammaphiOm}), may be applied. Then the identity (\ref{IIBi}) reads
\begin{equation}
d{{\raisebox{-0.1ex}{$I$}\!\!I}}[\Omega]_{AB} + 
2{\Gamma}[\Omega]^{K}{}_{(A}\wedge 
{{\raisebox{-0.1ex}{$I$}\!\!I}}[\Omega]_{B)K} + 
{{\cal B}}[\Omega]_{AB}=0.\label{IIBiid}
\end{equation}
We see that the fulfillment of (D) yields (C) of the theorem~11.
It is worth noting that the algebraic constraints (A) are always assumed 
to be true, which is necessary for the derivation of (\ref{IIBiid}) from 
(\ref{IIBi}).

Similarly, it follows from (\ref{BiII})
\begin{equation}
d{{\cal B}}[\Omega]_{AB}+
2{\Gamma}[\Omega]^{K}{}_{(A}\wedge {{\cal B}}[\Omega]_{B)K}=
2{{\raisebox{-0.1ex}{$I$}\!\!I}}[\Omega]^{K}{}_{(A}\wedge
{\Omega}_{B)K},\label{BiIIid}
\end{equation}
In contrast to (\ref{IIBiid}), the above equation does not imply
${{\raisebox{-0.1ex}{$I$}\!\!I}[\Omega]}_{AB}=0$ whenever the Bianchi 
equations ${{\cal B}}[\Omega]_{AB}=0$ are satisfied, but rather a less 
restrictive constraint ${{\raisebox{-0.1ex}{$I$}\!\!I}[\Omega]}^{K}{}_{(A}
\wedge{\Omega}_{B)K}=0$ is entailed. Nevertheless 
some equations from the closed system become linear dependent.

Further, the way of introduction of the operator $\Phi[\Omega]$ implies 
the additional identity
\begin{equation}
{\widetilde C}^{ABCD}\ \widetilde{\ast}
({{\cal B}}[\Omega]_{AB})\wedge{\Omega}_{CD}=0,\label{phiid1}
\end{equation}
which in its turn entails, by virtue of (\ref{IIBiid}),
the closed linear homogeneous equation restricting possible values of 
${{\raisebox{-0.1ex}{$I$}\!\!I}[\Omega]}_{AB}$ with $\Omega$'s obeying 
algebraic constraints (A):
\begin{equation}
{\widetilde C}^{ABCD}\ \widetilde{\ast}(d{{\raisebox{-0.1ex}{$I$}\!\!I}}
[\Omega]_{AB} + 
2{\Gamma}[\Omega]^{K}{}_{A}\wedge {{\raisebox{-0.1ex}{$I$}\!\!I}}
[\Omega]_{BK})\wedge{\Omega}_{CD}=0.
\label{phiid2}
\end{equation}

A consequence of the last identity (\ref{IIS}) is a little subtle. In 
contrast to (\ref{BiII}), (\ref{IIBi}) which are true for arbitrary
$\Gamma,\Omega$, the eq.~(\ref{IIS}) is valid only if the first structure
equations (\ref{Feq}) are fulfilled. In the framework of the current 
section, the connection (\ref{explGamma}) obeys the structure equations 
(\ref{Feq}) automatically but the expression (\ref{GammaphiOm}) {\it does
not}, in general. However it is easy to see that ${\Gamma[\Omega]}_{AB}$ 
still obeys (\ref{Feq}) provided $\Phi[\Omega]=d\phi$ for some (arbitrary)
function $\phi$ and $S_{AB}=e^{\phi}\,{\widetilde S}_{AB}$, the latter 
relation being regarded here as the definition of its l.h.s. In such a 
case therefore ${{\raisebox{-0.1ex}{$I$}\!\!I}[\Omega]}^{K}{}_{(A}\wedge
{\widetilde S}_{B)K}=0$
for {\it arbitrary} $\phi$. On the other hand, from the formal point of 
view this equation could fail in general case, that is after the 
replacement $d\phi \rightarrow \Phi[\Omega]$, only because then 
$d\Phi[\Omega]$, unlike $dd{\phi}=0$, does not vanish identically. Thus 
the restoring of $\Phi[\Omega]$ in place of $d\phi$ in the adapted version 
of the identity (\ref{IIS}) introduces the additional term proportional to
$d\Phi[\Omega]$. The latter can easily be calculated yielding the final 
identity
\begin{equation}
-3 d\Phi[\Omega]\wedge\widetilde{S}_{AB}+
2{{\raisebox{-0.1ex}{$I$}\!\!I}[\Omega]}^{K}{}_{(A}\wedge
{\widetilde S}_{B)K}=0.\label{phiIIid}
\end{equation}
Note that if the structure equations ${{\raisebox{-0.1ex}{$I$}\!\!I}
[\Omega]}_{AB}=0$ are fulfilled then $\Phi[\Omega]$ obeys the 
complexified source-free Maxwell-type equations.

We have proven
\medskip

{\bf Proposition 12}\ {\it 
If the algebraic constraints (A) of theorem 11 are fulfilled
then the l.h.s.'s of the equations (B)-(D) obey a series of identities
presented by} (\ref{IIBiid})--(\ref{phiIIid}).

\section{Discussion}

The physical importance of  space-time curvature which distinguishes
the gravitational and  purely inertial effects suggests that the 
components of the curvature, rather than that of the metric or connection,
should be interpreted as the mathematical representative of the ``true" 
gravitational field. Then the problem arises to describe the main physical 
and geometrical structures in terms of this fundamental object. The first 
tractable case of a significant physical interest is the one of the 
absence of extended sources of gravity, i.e. the case of a vacuum 
space-time which  was  considered above.

In the case of a non-zero cubic conformal curvature invariant, a generic
vacuum Einstein space-time curvature is endowed with an extremely simple 
algebraic structure. The crucial point is the use of a special family of 
2-forms, named above $S$-forms, which span the subspace of anti-self-dual 
elements of the complexified 2-forms space and are fixed up to  
$SO(3,C)$ group transformations. A remarkably simple 
quartic constraint imposed on the components of the (anti-self-dual 
complex of) curvature 2-forms (see item (A) of the theorem 11) is 
necessary for the latter to be associated with a  vacuum geometry. If 
it is fulfilled, the metric can be restored from the curvature 
components up to a conformal factor (see the eq.~(\ref{gphiOm})),
with the conformal metric components being  homogeneous rational 
functions of the components of curvature 2-forms.

Further, the conformal factor $\phi$, which is necessary for the complete
determination of the geometry, is calculated from the equation 
$d\phi=\Phi[\Omega]$ formulated in terms of  curvature components.
The latter equation is automatically consistent due to the Bianchi
identities.

Keeping these basic relations in mind, it is straightforward to deduce
a complete set of the vacuum gravitational field equations (including the
Bianchi equations) in a closed form in terms of the curvature components
alone. These are listed in  theorem~11. An important intermediate
step, which seems to be worthwhile mentioning, is the derivation of an
explicit closed representation for the connection 1-forms in terms of the
curvature components (eq.~(\ref{GammaphiOm})).

The auxiliary metric $\widetilde{g}$ with the relevant Hodge operator
$\widetilde{\ast}$ turn out to be a convenient {\it technical} tool
which enables the complete reformulation of the vacuum Einstein theory
in terms of the curvature. Of more fundamental importance is the general
formalism of $S$-forms which is very helpful, in particular, in 
discussing  exact solutions of the  gravitational field equations. 
The present paper contains the formal general framework. Its applications
to the study of exact solutions will be considered elsewhere.

\bigskip
{\bf Acknowledgments}. 
\bigskip

We would like to thank Friedrich W. Hehl for very useful criticism and
advice and A.A. Tseytlin for careful reading of the manuscript. 
The work of  YNO  was supported by the Deutsche 
Forschungsgemeinschaft (Bonn) grant He 528/17-1.

\end{document}